\documentclass[10pt,aps,prb]{revtex4}
\usepackage{amssymb}
\usepackage{amsfonts}

\begin{document}

\title{Application of exterior calculus to waveguides}
\author{Rafael Ferraro}
\affiliation{Instituto de Astronom\'\i a y F\'\i sica del Espacio,
Casilla de Correo 67, Sucursal 28, 1428 Buenos Aires, Argentina}
\affiliation{Departamento de F\'\i sica, Facultad de Ciencias
Exactas y Naturales, Universidad de Buenos Aires, Ciudad
Universitaria, Pabell\'on I, 1428 Buenos Aires, Argentina}
\email{ferraro@iafe.uba.ar}
\thanks{Member of Carrera del Investigador Cient\'{\i}fico (CONICET,
Argentina)}

\begin{abstract}
Exterior calculus is a powerful tool to search for solutions to the
electromagnetic field equations, whose strength can be better appreciated
when applied to work out non-trivial configurations. Here we show how to
exploit this machinery to obtain the electromagnetic TM and TE modes in
hollow cylindrical waveguides. The proper use of exterior calculus and
Lorentz boosts will straightforwardly lead to such solutions and the
respective power transmitted along the waveguide.
\end{abstract}

\maketitle

%\pacs{Valid PACS appear here} \keywords{exterior calculus, differential forms, waveguides}

\section{Introduction}

Waveguides are an excellent arena to practice relativity and
calculus with differential forms, since these tools greatly help
the understanding of electromagnetic problems and their solutions.
Although most of textbooks teaching exterior calculus to
physicists display some basic applications to
electromagnetism,\cite{schu,flan,darl,west} any of them stresses
the power of this language by applying it to solve Maxwell
equations in waveguides and cavities. This is rather
disappointing, because the skill in applying both tools
--relativity and exterior calculus-- allows a clean presentation
of the subject and an easy way for working out the field
configurations, so highlighting the power of the language not only
as a theoretical weapon but as a practical tool to solve problems
of technical relevance. On the contrary, normally the vector
language is used to display the field configurations in waveguides
(however see Ref.~\onlinecite{russ}), which is hardly a natural
language for electromagnetism. Because of this reason, the vector
approach tends to be rather tedious. Frequently, the lack of a
proper geometric language reduces the explanation to a simple case
like the waveguide of rectangular or circular section (however see
Ref.~\onlinecite{jack}). We will work out the problem of
propagating waves in hollow cylindrical waveguides of arbitrary
section by employing exterior calculus. We will start with TM (TE)
(non-propagating) stationary modes, whose field structure is very
simple: it is just an electric (magnetic) field along the
waveguide and a magnetic (electric) field in the waveguide
section. Then, we will turn these stationary solutions into
propagating solutions by means of a Lorentz boost along the
waveguide. Finally we will compute the transmitted power and
emphasize its relativistic relation with the energy per unit of
length in the waveguide. The calculus involving exterior
derivatives, Hodge dualities and the generalized Stokes theorem
will provide a straightforward way for building the solutions,
since all the vector equations of the usual approach condense in
just one equation in geometric language, so showing how powerful
this machinery is.

\bigskip

\section{A brief review of exterior calculus}

Exterior calculus is the natural language for
electrodynamics.\cite{bamb,curt} But not only electrodynamics
greatly benefits from the compactness and simplicity of this
geometric language. Also the developments in Hamiltonian
mechanics,\cite{arno,west} thermodynamics,\cite{schu,west}
Yang-Mills fields,\cite{baez,naka} geometric (Berry) phases in
quantum mechanics,\cite{fran} topological quantum fields as the
Chern-Simons theory,\cite{baez,naka} gravity,\cite{misn}
symplectic geometry,\cite{guil} connections in fiber
bundles,\cite{koba} etc. gain in clarity and depth when expressed
through the tools of exterior calculus. There is a huge list of
textbooks to learn exterior calculus. The reader is referred to
Ref.~\onlinecite{schu,flan,darl,west} for a first approach to the
subject. Here we will briefly review the main operational features
of the exterior derivative $d$ and the wedge product $\wedge$
between differential forms.

Any linear combination of coordinate differentials at each point of space is
a \emph{1-form} field (whatever the coordinates are, Cartesian or not; even
if the geometry is non-Euclidean). For instance, we may call $\eta $ the
1-form
\begin{equation}
\eta =3\, x^{2}y^{7}\,dx+5\,y\;dz
\end{equation}
If the space is 3-dimensional and $(x,y,z)$ are the chosen
coordinates, we say that $\{dx,dy,dz\}$ is a coordinate basis for
1-forms. The \emph{components} of the above defined 1-form $\eta$
in this basis are $\eta _{x}= $ $3\, x^{2}y^{7}$, $\eta _{y}=$
$0$, $\eta _{z}=$ $5\,y$. Generically, a 1-form in a
$n$-dimensional space\ is
\begin{equation}
\alpha =\alpha _{\mu }\;dx^{\mu }  \label{eq00}
\end{equation}
(Einstein convention is used). The superindex in $dx^{\mu }$
labels the $n$ 1-forms of the coordinate basis, whereas the
subindex in $\alpha _{\mu }$ labels the components of the 1-form
$\alpha $, which are functions of the coordinates.

1-forms can be introduced in a geometric way as linear real valued
functions on the tangent vector space to a manifold: they are
covectors or covariant vectors.\cite{schu,west} However, here we
are not interested in the action of forms on vectors. Instead we
will operate within the set of \emph{p-forms}, which are defined
as totally antisymmetric covariant tensors of $p$ indexes (so
$p\leqq n$). $p$-forms can be obtained from (antisymmetrized)
\emph{wedge }tensor product $\wedge $\ of 1-forms. For instance,
the wedge product between $\eta$ and the 1-form $\xi =z\;dx+2\;dy$
is the 2-form
\begin{equation}
\varpi =\eta \wedge \xi =6\, x^{2}y^{7}\,dx\wedge
dy-5\,y\,z\;dx\wedge dz-10\, y\;dy\wedge dz
\end{equation}
Notice the absence of $dx\wedge dx\equiv 0$, since the wedge
product is antisymmetric; we also use that $dx\wedge dz=-dz\wedge
dx$. There are ${n\choose 2}$ linearly independent 2-forms
$dx^{\mu }\wedge dx^{\nu }\equiv dx^{\mu }\otimes dx^{\nu
}-dx^{\nu }\otimes dx^{\mu }$ ($\otimes $ is the tensor product),
which span the coordinate basis of 2-forms. Any 2-form can be
written as
\begin{equation}
\alpha =\frac{1}{2!}\;\alpha _{\mu \nu }\;dx^{\mu }\wedge dx^{\nu }
\label{eq4}
\end{equation}
where $\alpha _{\mu \nu }=-\alpha _{\nu \mu }$ . In the former
example it is $\varpi _{xy}=$ $6\, x^{2}y^{7}$, $\varpi _{yz}=$
$0$, $\varpi _{xz}=-5\,y\,z$, The factor $1/2!$ in Eq.~(\ref{eq4})
takes into account the fact that each independent element of the
basis appears $2!$ times in the sum over $\mu $ and $\nu $.

If $\alpha $ and $\beta $ are 1-forms on a 3-dimensional manifold,
then the components of the product $\alpha \wedge \beta $ will
look as the Cartesian components of the vector product in
Euclidean space:
\begin{equation}
\alpha \wedge \beta =(\alpha _{x}\beta _{y}-\alpha _{y}\beta _{x})\,dx\wedge
dy-(\alpha _{x}\beta _{z}-\alpha _{z}\beta _{x})\,dz\wedge dx+(\alpha
_{y}\beta _{z}-\alpha _{z}\beta _{y})\,dy\wedge dz  \label{eq01}
\end{equation}

Any $p$-form $\alpha $ and $q$-form $\beta$ satisfy
\begin{equation}
\alpha \wedge \beta =(-1)^{pq}\;\beta \wedge \alpha  \label{eq02}
\end{equation}
Thus 1-forms anti-commute but 2-forms commute, etc.

The exterior derivative $d$ is a nilpotent operator ($d^{2}\equiv
0$). If $d$ acts on a function $f$ (0-form) the result is the
1-form $df=(\partial f/\partial x^{\mu })\,dx^{\mu }$. In general,
if $d$ acts on a $p$-form $\alpha $ then the result is a
($p+1$)-form $d\alpha $. Since $d(dx^{\mu })\equiv 0$, $d\alpha$
is obtained by merely differentiating its components as exterior
derivatives of functions:
\begin{equation}
d\alpha =d\left( \frac{1}{\,p!}\;\alpha _{\lambda \mu \nu ....}\right)
\wedge dx^{\lambda }\wedge dx^{\mu }\wedge dx^{\nu }....  \label{eq03}
\end{equation}
For instance
\begin{equation}
d\eta =-21\,x^{2}y^{6}\,dx\wedge dy+5\;dy\wedge dz
\end{equation}
If $\alpha $ is a $p$-form then
\begin{equation}
d(\alpha \wedge \beta )=d\alpha \wedge \beta +(-1)^{p}\;\alpha \wedge d\beta
\label{eq2}
\end{equation}
A given $p$-form $\alpha $ is called \emph{closed} if $d\alpha
=0$; it is called \emph{exact} if it can be written as the
exterior derivative of a ($p-1$)-form. Each exact form is closed;
the inverse proposition is locally true, but its global validity
depends on the topology of the space (see Poincar\'{e}'s lemma
\cite{schu,west,fran}).

Finally, we will introduce the Hodge star operator $\ast $. This
operator changes $p$-forms into $(n-p)$-forms, and involves the
components $g_{\mu \nu }$ of the metric tensor present in the
space-time interval. If $\alpha $ is a $p$-form whose components
are $\alpha _{\mu _{1}.......\mu _{p}}$ then
\begin{equation}
\ast \alpha _{\mu _{p+1}.......\mu _{n}}=\frac{1}{\,p!}\,\sqrt{|\det
(g_{\mu \nu })|} \ \varepsilon _{\mu _{1}.......\mu _{p}\;\mu
_{p+1}.......\mu _{n}}\;\alpha ^{\mu _{1}.......\mu _{p}}
\label{eq7}
\end{equation}
where the indexes are raised with the inverse metric tensor
$g^{\mu \nu }$: $\alpha ^{\mu _{1}.......\mu _{p}}=g^{\mu _{1}\nu
_{1}}.....g^{\mu _{p}\nu _{p}}\;\alpha _{\nu _{1}.......\nu
_{p}}$. $\varepsilon $ is the Levi-Civita symbol, which takes the
value $1$ ($-1$) for even (odd) permutations of the natural order
of its indexes, and it is zero if there are indexes of equal
value. The successive application of the Hodge star operator on a
$p$-form $\alpha $ is \cite{flan,darl}
\begin{equation}
\ast \ast \alpha =(-1)^{p(n-p)+(n-\sigma )/2}\ \alpha  \label{eq40}
\end{equation}
where $\sigma $ is the signature of the metric tensor (the
difference between the numbers of positive and negative
eigenvalues of the metric tensor).

\section{Potential and field as differential forms on a manifold}

The electromagnetic field is an exact 2-form $F=dA$, where the
1-form $A$ is the potential. Given a set of non-necessarily
Cartesian coordinates $x$, $y$, $z$ together with the time $t$,
one can express these forms in the coordinate basis
$\{dt,dx,dy,dz\}$:
\begin{equation}
A=A_{\nu }\;dx^{\nu }  \label{eq1}
\end{equation}
If $x$, $y$, $z$ are Cartesian coordinates, then the components
$A_{\nu }$ will coincide with the scalar and vector potential:
$A_{\nu }=(-\phi ,\mathbf{A})$ (SI units) or $A_{\nu }=(-c\,\phi
,\mathbf{A})$ (Gaussian units). Therefore
\begin{equation}
F=dA=dA_{\nu }\wedge dx^{\nu }=\partial _{\mu }A_{\nu }\;dx^{\mu }\wedge
dx^{\nu }  \label{eq3}
\end{equation}
Since $dx^{\mu }\wedge dx^{\nu }=dx^{\mu }\otimes dx^{\nu
}-dx^{\nu }\otimes dx^{\mu }$, then the (antisymmetric) components
of $F$ are $F_{\mu \nu }=\partial _{\mu }A_{\nu }-\partial _{\nu
}A_{\mu }$. For Cartesian coordinates, it is $F_{i\,t}=E_{i}$ in
SI units (or $cE_{i}$ in Gaussian units), $F_{\,yz}=B_{x}$,
$F_{zx}=B_{y}$, $F_{\,xy}=B_{z}$:
\begin{equation}
F=\ E\wedge dt+B_{x}\ dy\wedge dz+B_{y}\ dz\wedge dx+B_{z}\ dx\wedge dy
\label{eq5}
\end{equation}
where $E=E_{i}\;dx^{i}$\ (SI units) is a 1-form.\ As Eq.
(\ref{eq5}) shows, in the 3-dimensional Euclidean space the wedge
product between 1-forms works as the vector product, since
$dy\wedge dz$ is the basis element for the $x$ -component of the
pseudo-vector $\mathbf{B}$, and so on (think the wedge product as
a vector product; see Eq.~(\ref{eq01})).

Since $F=dA$ is exact and $d$ is nilpotent, then it results the identity
\begin{equation}
dF=0  \label{eq6}
\end{equation}
which amounts those Maxwell equations used to define the
potentials: $ \mathbf{\nabla }\times \mathbf{E}=-\partial
\mathbf{B}/\partial t$, $\mathbf{\nabla }\cdot \mathbf{B}=0$.

The dynamical sourceless Maxwell equations, $\mathbf{\nabla
}\times \mathbf{B }=-c^{-2}\ \partial \mathbf{E}/\partial t$,
$\mathbf{\nabla }\cdot \mathbf{E}=0$, come from varying the
electromagnetic action $S[A]=-(4\mu _{o}c)^{-1}\int \ast F\wedge
F$. Since $F$ is a 2-form in a manifold of dimension $n=4$ (the
spacetime), then $\ast F$ is a 2-form too. $\ast F\wedge F$ is a
4-form (a \textsl{volume} in spacetime). The resulting Euler-
Lagrange equations are
\begin{equation}
d\ast F=0  \label{eq8}
\end{equation}

\section{Cylindrical waveguides}

We are going to solve the sourceless Maxwell equations for the
electromagnetic field in hollow cylindrical waveguides. So, let us
call $z$ the Cartesian coordinate along the waveguide, and $x$,
$y$ two properly chosen coordinates spanning the waveguide
section: Cartesian coordinates for rectangular section, polar
coordinates for circular section, etc. We will begin by working
out stationary waves. Afterwards we will introduce the propagation
along the waveguide by means of a Lorentz boost in the $z$
-direction. Therefore we will start by proposing a solution
independent of $z$.

\subsection{Stationary TM modes}

Let us consider a monochromatic potential having just a component
along the waveguide:
\begin{equation}
A=e^{i\Omega t}\;\psi (x,y)\;dz  \label{eq9}
\end{equation}
Function $\psi $ is defined in the waveguide section and has units of
magnetic field times length. Thus the electromagnetic field is
\begin{eqnarray}
F &=&dA=e^{i\Omega t}\;\left( i\Omega \;\psi \ dt+\ d\psi \right) \wedge dz
\nonumber \\
&=&e^{i\Omega t}\;\left( i\Omega \;\psi \ dt\wedge dz+\ \partial _{x}\psi
\;dx\wedge dz+\ \partial _{y}\psi \;dy\wedge dz\right)  \label{eq10}
\end{eqnarray}
The first term is an electric field along $z$ and the other ones
make up a magnetic field orthogonal to the $z$-axis, so lying on
the waveguide section. Therefore the proposed solution is a TM
mode. For the moment this solution does not propagate along the
waveguide since the components do not depend on $z$ (this feature
will be introduced later). Field (\ref{eq10}) is just a stationary
wave bouncing between the boundaries. Function $\psi $ is
subjected to fulfill perfect conductor boundary conditions: the
tangential electric field and the normal magnetic field must
vanish on the boundary. Then $\psi $ must be zero on the boundary
to cancel out the (pure) tangential electric field:
\begin{equation}
\psi |_{boundary}=0  \label{eq11}
\end{equation}
(Dirichlet condition for the potential $\psi (x,y)$). Since $\psi
$ is constant on the boundary then the 1-form $d\psi =(\partial
\psi /\partial x^{\mu })\,dx^{\mu }$ is \emph{normal} to the
boundary; so the magnetic part $d\psi \wedge dz$ is tangential to
the boundary and perpendicular to the $z$ -axis (just think the
product as a vector product). Thus, the boundary condition for the
electric field guarantees the boundary condition for the magnetic
field too.

The 2-form $\ast F$ involves the metric tensor. If orthogonal
coordinates are used in the waveguide section, then the space-time
interval will read
\begin{equation}
ds^{2}=-c^{2}dt^{2}+g_{xx}\ dx^{2}+g_{yy}\ dy^{2}+dz^{2}  \label{eq12}
\end{equation}
where the metric \textrm{diag}$(g_{xx},g_{yy})$ in the section
depends on the choice of coordinates $x$, $y$. Then the
determinant of the metric and the inverse metric tensor are
\begin{equation}
|\det (g_{\mu \nu })|=c^{2}\;g_{xx}\;g_{yy}\;,\ \ \ \ \ \ \ \ \ \
\ \ g^{\mu \nu }=\text{diag}(-c^{-2},1/g_{xx},1/g_{yy},1)
\label{eq13}
\end{equation}
To write $\ast F$ we just need a few results:
\begin{eqnarray}
\ast (dt\wedge dz) &=&-c^{-1}\,\sqrt{g_{xx}g_{yy}}\ dx\wedge dy\;,\
\nonumber \\
\ast (dx\wedge dz) &=&-c\,\sqrt{g_{xx}g_{yy}}\ g^{xx}\ dt\wedge dy\;,\
\label{eq14} \\
\ast (dy\wedge dz) &=&c\,\sqrt{g_{xx}g_{yy}}\ g^{yy}\ dt\wedge dx  \nonumber
\end{eqnarray}
Thus,
\begin{equation}
\ast F=c\,\sqrt{g_{xx}g_{yy}}\ e^{i\Omega t}\ \left\{ -ic^{-2}\Omega \;\psi
\ dx\wedge dy+\left( \ g^{xx}\;\partial _{x}\psi \;dy-\ g^{yy}\;\partial
_{y}\psi \;dx\right) \wedge dt\right\}  \label{eq15}
\end{equation}
\begin{equation}
d\ast F=c\sqrt{g_{xx}g_{yy}}\;e^{i\Omega t}\;\left( c^{-2}\Omega ^{2}\;\psi
+^{(2)}\!\!\Delta \psi \right) \;\ dt\wedge dx\wedge dy  \label{eq16}
\end{equation}
Eq.~(\ref{eq8}) implies that function $\psi $ must be an
eigenfunction $\psi _{mn}$ of the two-dimensional Laplacian
operator in the waveguide section, $-c^{-2}\Omega_{mn}^{2}$ being
its respective eigenvalue:
\begin{equation}
^{(2)}\!\!\Delta \psi _{mn} \equiv
\frac{1}{\sqrt{g_{xx}g_{yy}}}\left[ \partial _{x}\
(\sqrt{g_{xx}g_{yy}}\;g^{xx}\;\partial _{x}\psi _{mn} )+\partial
_{y}(\sqrt{ g_{xx}g_{yy}}\;g^{yy}\ \partial _{y}\psi _{mn}
)\right] =-\frac{\Omega_{mn}^{2}}{c^{2}} \;\psi _{mn}
\label{eq17}
\end{equation}
Notice that Eq.~(\ref{eq15}) contains the two-dimensional 1-form
$^{(2)}\!\!(\ast d\psi )$ (the superscript \textquotedblleft $(2)$
\textquotedblright\ means that the Hodge star is applied in a
$n=2$ submanifold; see Appendix). As can be seen, the Laplacian in
the waveguide section is $^{(2)}\!\!\Delta =^{(2)}\!\!(-\ast d\ast
d)$. Its eigenfunctions $\psi _{mn}$ satisfying proper boundary
conditions are identified by two discrete indexes $m,n$. Table
\ref{table1} summarizes the results for monochromatic TM
stationary modes. Table \ref{table2} shows the solutions $\psi$ of
Eqs.~(\ref{eq11}), (\ref{eq17}) and their eigenvalues $\Omega $
(allowed frequencies for stationary TM modes) for typical
waveguide sections.

\begin{table}[h]
\caption{TM stationary modes.} \label{table1}
\begin{tabular}{|c|}
\hline\hline
\\
$F^{TM}=e^{i\Omega_{mn} t}\;\left( i\,\Omega _{mn}\;\psi _{mn}\,\ dt+\ d\psi
_{mn}\,\right) \wedge dz$ \\
\\
$\ \ \ \ast F^{TM}=e^{i\Omega_{mn} t}\;\left( -i\,c^{-1}
\Omega_{mn}\,\sqrt{g_{xx}g_{yy}} \ \psi_{mn}\,\ dx\wedge
dy+c\,^{(2)}\!\!(\ast d\psi_{mn})\wedge dt\right)\ \ \ $ \\
\\
\textbf{$\psi$ and $^{(2)}\!\!(\ast d\psi)$ are defined in the waveguide
section:} \\[1ex]
$^{(2)}\!\!\Delta
\psi_{mn}(x,y)\,=-c^{-2}\,\Omega_{mn}^2\,\psi_{mn}(x,y)\,\,\ \ \ \ \ \ \ \ \
\ \ \ \ \ \psi_{mn}\,|_{boundary}=0$ \\
\\
\textbf{metric in the section:}\ \ \ \ \ $d\ell^2=g_{xx}(x,y)
\;dx^2+g_{yy}(x,y)\;dy^2$ \\
\\ \hline
\end{tabular}
\end{table}

\begin{table}[h]
\caption{TM $\protect\psi$'s and $\Omega$'s for rectangular and
circular sections.} \label{table2}
\begin{tabular}{|c|}
\hline\hline \textbf{Rectangular section} \\ $x,y$ are Cartesian
coordinates, $0\leq x\leq a$, $0\leq y\leq b$, $g_{xx}=1$,
$g_{yy}=1$ \\[0.8ex] $\psi _{mn}(x,y)=A_{mn}\sin \left( \frac{m\pi
}{a}x\right) \sin \left( \frac{n\pi}{b}y\right) \;,\ \ \ \
m,n\in\mathbb{N}$ \\[0.8ex] $c^{-2}\Omega _{mn}^{2}=\left(
\frac{m\pi }{a}\right) ^{2}+\left( \frac{n\pi }{b}\right) ^{2}$ \\
\\
\textbf{Circular section} \\[0.8ex] $\ \ \ \ \ x,y$ are polar
coordinates $r, \varphi$: $0\leq r\leq R$, $0\leq \varphi \leq
2\pi$, $g_{rr}=1$, $g_{\varphi \varphi }=r^{2}\ \ \ \ \ $ \\
[0.8ex] $\psi _{mn}(r,\varphi )=J_{m}(x_{mn}\,r\,/R)\;\left(
A_{mn}\;\cos (m\varphi )+B_{mn}\;\sin (m\varphi )\right)$
\\[0.8ex] $\Omega _{mn}=\frac{x_{mn}\;c}{R}$ \ \ \ \ \ \ \ \
$x_{mn}$ are zeros of Bessel functions $J_{m}$ \\
\\ \hline
\end{tabular}
\end{table}

\subsection{Stationary TE modes}

\bigskip Remarkably $F$ and $\ast F$ are on an equal footing in
Eqs.~(\ref{eq6}) and (\ref{eq8}). This circumstance will allow to
built the stationary TE modes by exchanging their roles. In Eq.
(\ref{eq5}), this exchange amounts to the interchanging of
$\mathbf{E}$ and $\mathbf{B}$. So we define
\begin{equation}
F^{TE}=\ast F^{TM}  \label{eq39}
\end{equation}
We will apply Eq.~(\ref{eq40}) to obtain $\ast F^{TE}=\ast \ast
F^{TM}$. In the space-time it is $n=4$ and $\sigma =2$ (see Eq.
(\ref{eq12})); so $\ast \ast =(-1)^{p+1}$. Thus
\begin{equation}
\ast F^{TE}=-F^{TM}
\end{equation}
The field (\ref{eq39}) can be ascribed to the potential
\begin{equation}
A^{TE}=\frac{e^{i\Omega t}c}{i\Omega }\sqrt{g_{xx}g_{yy}}\;\left(
g^{yy}\;\partial _{y}\psi \;dx\ -g^{xx}\;\partial _{x}\psi
\;dy\right) = \frac{e^{i\Omega t}c}{i\Omega }\;^{(2)}\!\!(\ast
d\psi )  \label{eq42}
\end{equation}
In fact, by differentiating the middle expression in Eq.
(\ref{eq42}) one recognizes the appearance of the Laplacian
defined in Eq.~(\ref{eq17}):
\begin{eqnarray}
\nonumber F^{TE}=dA^{TE}&=&e^{i\Omega t}\,c\,\sqrt{
g_{xx}g_{yy}}\,\left\{ \frac{i}{\Omega }\,\ ^{(2)}\!\!\Delta \psi
\;dx\wedge dy+\left( \ g^{xx}\;\partial _{x}\psi \;dy-\
g^{yy}\;\partial _{y}\psi \;dx\right) \wedge dt\right\}\\
&=&e^{i\Omega t}\,c\,\sqrt{ g_{xx}g_{yy}}\,\left\{ \frac{i}{\Omega
}\,\ ^{(2)}\!\!\Delta \psi \;dx\wedge dy+ ^{(2)}\!\!(\ast d\psi
)\wedge dt\right\} \label{eq41}
\end{eqnarray}
This result coincides with field (\ref{eq15}) when the pair
$\Omega, \psi$ is chosen among the solutions $\Omega_{mn},
\psi_{mn}$ of the eigenvalue equation (\ref{eq17}). Although
$F^{TE}$ solves both Eqs.~(\ref{eq6}) and (\ref{eq8}), the
boundary condition should be consistently changed: the electric
field should be normal to the boundary to accomplish the perfect
conductor boundary condition. If $n=n_{x}\,dx+n_{y}\,dy$ is a
1-form normal to the boundary, and $^{(2)}\!\!(\ast d\psi )$ in
Eq.~(\ref{eq41}) is a 1-form proportional to the electric field in
the waveguide section, then one demands
\begin{equation}
^{(2)}\!\!(\ast d\psi )\wedge n\;|_{boundary}=0\label{eq62}
\end{equation}
(just think the product as a vector product). When written in vector
language this requirement means
\begin{equation}
\mathbf{n}\cdot \mathbf{\nabla }\psi \;|_{boundary}=0
\end{equation}
(Neumann boundary condition for the potential $\psi (x,y)$). Table
\ref{table3} summarizes the monochromatic stationary TE modes.
Table \ref{table4} shows the functions $\psi $ for typical
waveguide sections.
\begin{table}[h]
\caption{TE stationary modes.} \label{table3}
\begin{tabular}{|c|}
\hline\hline
\\
$\ \ \ F^{TE}=e^{i\Omega_{mn} t}\;\left(
-i\,c^{-1}\Omega_{mn}\,\sqrt{g_{xx}g_{yy}} \ \psi _{mn}\,\
dx\wedge dy+c\;^{(2)}\!\!(\ast d\psi _{mn})\wedge dt\right)\ \ \ $
\\
\\
$\ast F^{TE}=-e^{i\Omega_{mn} t}\;\left(
i\,\Omega_{mn}\;\psi_{mn}\,\ dt+\ d\psi_{mn}\,\right) \wedge dz$
\\
\\
\textbf{$\psi$ and $^{(2)}\!\!(\ast d\psi)$ are defined in the
waveguide section:} \\[1ex] $^{(2)}\!\!\Delta
\psi_{mn}(x,y)\,=-c^{-2}\,\Omega_{mn}^2\,\psi_{mn}(x,y)\,\,\ \ \ \
\ \ \ \ \ \ \ \ \ \ \mathbf{n}\cdot \mathbf{\nabla }\psi
\;|_{boundary}=0$ \\
\\
\textbf{metric in the section:}\ \ \ \ \ $d\ell^2=g_{xx}(x,y)
\;dx^2+g_{yy}(x,y)\;dy^2$ \\
\\ \hline
\end{tabular}
\end{table}

\begin{table}[h]
\caption{TE $\protect\psi$'s and $\Omega$'s for rectangular and
circular sections.} \label{table4}
\begin{tabular}{|c|}
\hline\hline \textbf{Rectangular section} \\ $x,y$ are Cartesian
coordinates, $0\leq x\leq a$, $0\leq y\leq b$, $g_{xx}=1$ ,
$g_{yy}=1$ \\[0.8ex] $\psi _{mn}(x,y)=A_{mn}\cos \left( \frac{m\pi
}{a}x\right) \cos \left( \frac{ n\pi }{b}y\right)$ \\[0.8ex]
$c^{-2}\Omega _{mn}^{2}=\left( \frac{m\pi }{a}\right) ^{2}+\left(
\frac{n\pi }{b}\right) ^{2}$ \\
\\
\textbf{Circular section} \\[0.8ex] $x,y$ are polar coordinates
$r, \varphi$: $0\leq r\leq R$, $0\leq \varphi \leq 2\pi$,
$g_{rr}=1$, $g_{\varphi \varphi }=r^{2}$ \\[0.8ex] $\psi
_{mn}(r,\varphi )=J_{m}(y_{mn}\,r\,/R)\;\left( A_{mn\;}\cos
(m\varphi )+B_{mn\;}\sin (m\varphi )\right)$ \\[0.8ex] $\ \ \ \ \
\Omega _{mn}=\frac{y_{mn}\;c}{R}$ \ \ \ \ \ \ \ \ $y_{mn}$ are
zeros of the derivatives of Bessel functions $J_{m}\ \ \ \ \ $ \\
\\ \hline
\end{tabular}
\end{table}

\section{Propagating modes}

The solutions studied in Section IV do not propagate energy along
the waveguide. Since the Poynting vector is proportional to
$\mathbf{E}\times \mathbf{B}$, then there should exist
$\mathbf{E}$ and $\mathbf{B}$ in the waveguide section to have
energy propagating along the waveguide. However, field $F^{TM}$ in
Eq.~(\ref{eq10}) has only a magnetic field in the waveguide
section, and field $F^{TE}$ in Eq.~(\ref{eq41}) has only an
electric field in the section. Thus, the Poynting vector in
solutions (\ref{eq10}), (\ref{eq41}) is orthogonal to the
waveguide axis;\ so solutions (\ref{eq10}), (\ref{eq41}) are
stationary waves where the energy bounces between the boundaries
but does not propagate along the waveguide. Moreover, it is easy
to prove that the time averaged Poynting vector vanishes in this
case. This means that the fields (\ref{eq10}), (\ref{eq41}) are
displayed in their \emph{proper frame}. The stationary solutions
(\ref{eq10}), (\ref{eq41}) can be transformed into solutions that
propagates energy along the waveguide by performing a Lorentz
boost in the $z$-direction. To do this we use
\begin{equation}
t=\gamma (V)(t^{\prime }-Vc^{-2}z^{\prime }),\ \ \ \ \ \ \ dt=\gamma
(V)(dt^{\prime }-Vc^{-2}dz^{\prime }),\ \ \ \ \ \ \ dz=\gamma (V)(dz^{\prime
}-Vdt^{\prime })  \label{eq24}
\end{equation}
where $\gamma (V)=(1-V^{2}c^{-2})^{-1/2}$. Thus
\begin{equation}
dt\wedge dz=\gamma (V)^{2}(1-V^{2}c^{-2})\;dt^{\prime }\wedge dz^{\prime
}=dt^{\prime }\wedge dz^{\prime }  \label{eq25}
\end{equation}

\subsection{TM modes}

Eq.~(\ref{eq25}) means that the longitudinal electric field
remains invariant in Eq.~(\ref{eq10}). On the contrary the
transverse term $d\psi (x,y)\wedge dz$ changes to
\begin{equation}
d\psi (x,y)\wedge dz=\gamma (V)\;d\psi (x,y)\wedge (dz^{\prime }-Vdt^{\prime
})  \label{eq26}
\end{equation}
which amounts not only a change of the magnetic field in the
waveguide section but the appearance of a transverse electric
field. Of course this is nothing but the usual rules to transform
electric and magnetic fields (see for instance
Ref.~\onlinecite{ferr}). However the geometric language shows it
in a quite elegant way.

\subsection{TE modes}

In this case the Eq.~(\ref{eq25}) means that the longitudinal
magnetic field remains invariant in Eq.~(\ref{eq41}). Instead, the
Lorentz boost changes the electric transverse sector of $F^{TE}$
to
\begin{eqnarray*}
c\;^{(2)}\!\!(\ast d\psi _{mn})\wedge dt &=&c\;\gamma (V)\,^{(2)}\!\!(\ast
d\psi _{mn})\wedge (dt^{\prime }-Vc^{-2}dz^{\prime }) \\
&=&c\;\gamma (V)\,\sqrt{g_{xx}g_{yy}}\,(\ g^{yy}\;\partial _{y}\psi \;dx-\
g^{xx}\;\partial _{x}\psi \;dy)\wedge (dt^{\prime }-Vc^{-2}dz^{\prime })
\end{eqnarray*}
Therefore, not only the electric transverse field is changed by the boost,
but a magnetic field appears in the waveguide section in the new frame.

\bigskip

As a conclusion of this Section, in any frame differing from the proper
frame where the solutions (\ref{eq10}), (\ref{eq41}) were built, the
propagating TM and TE modes display both electric and magnetic fields in the
waveguide section and so they propagate energy along the waveguide.

\section{Transmitted power}

As already stated, the existence of both magnetic and electric
transverse fields in the new frame produces a Poynting vector
along the waveguide. Thus, in the new frame there is energy
propagating in the waveguide. Velocity $V$ is the velocity
relative to the proper frame. In this sense $V$ has the right to
be called \emph{energy velocity}, since no energy propagates along
the waveguide in the original proper frame (there is just an
energy flux orthogonal to the waveguide axis whose time averaging
vanishes).

The time averaged energy flux along the waveguide, in the frame
moving with velocity $V$ relative to the proper frame, results
from the $t^{\prime }z^{\prime }$ component of the electromagnetic
energy-momentum tensor $T_{\mu \nu }$ \cite{jack}
\begin{equation}
\mu _{o}\,T_{\mu \nu }=F_{\mu \rho }F_{\nu }^{\ \ \rho
}-\frac{1}{4} g_{\mu \nu }F_{\lambda \rho }F^{\lambda \rho }
\label{eq27}
\end{equation}
Only real fields should be considered; so one has to average
products of trigonometric functions. It is well known that $<\sin
^{2}(\Omega t)>=\frac{1}{2}=<\cos ^{2}(\Omega t)>$, $<\sin (\Omega
t)\cos (\Omega t)>=0$. Thus one obtains
\begin{equation}
\mu _{o}<T_{t^{\prime }z^{\prime }}>=g^{xx}<F_{t^{\prime }x}F_{z^{\prime
}x}>+g^{yy}<F_{t^{\prime }y}F_{z^{\prime }y}>=-\frac{1}{2}\,\gamma
(V)^{2}V\;|\mathbf{\nabla }\psi |^{2}  \label{eq28}
\end{equation}
(we remark that the Lorentz boost does not change the components
of the metric tensor). The result (\ref{eq28}) is shared by TM and
TE modes (although it is harder to get it for TE modes).

$c^{2}T^{t^{\prime }z^{\prime }}=-T_{t^{\prime }z^{\prime }}$ is
the energy per unit of time and area going through the waveguide
section. The transmitted power results from integrating this
quantity in the section $S$. Since $\psi$ vanishes on the boundary
of the section (its normal derivative vanishes at the boundary for
TE modes), then the Green's first identity implies that
\cite{footnote}
\begin{equation}
\int \mathbf{\nabla }\psi \cdot \mathbf{\nabla}\psi \ dS=-\int
\psi \;^{(2)}\!\!\Delta \psi \ dS
\end{equation}
(both for TM and for TE modes). Thus one uses Eq.~(\ref{eq17}) to
obtain
\begin{equation}
\int |\mathbf{\nabla }\psi |^{2}\ dS=\int c^{-2}\Omega ^{2}\;\psi
^{2}\ dS  \label{eq29}
\end{equation}
Therefore the transmitted power is
\begin{equation}
P_{mn}=\int c^{2}T^{t^{\prime }z^{\prime }}\,dS=\frac{\gamma
(V)^{2}V\,\Omega _{mn}^{2}}{2\,\mu _{o}\,c^{2}}\int \psi _{mn}^{2}\,dS
\label{eq30}
\end{equation}
This result can be written in terms of the frequency $\omega $ and the
wavenumber $k_{z^{\prime }}$. The transformation of the coordinate $t$
implies that the wave phase becomes
\begin{equation}
\Omega \,t=\Omega \,\gamma (V)(t^{\prime }-Vc^{-2}z^{\prime })  \label{eq31}
\end{equation}
Then
\begin{equation}
\omega =\gamma (V)\Omega ,\ \ \ \ \ \ k_{z^{\prime }}=\gamma (V)\Omega
Vc^{-2}\   \label{eq32}
\end{equation}
which leads to the dispersion relation
\begin{equation}
\omega =\sqrt{c^{2}\ k_{z^{\prime }}^{2}+\Omega ^{2}}  \label{eq33}
\end{equation}
(then $\Omega _{mn}$ is the cut-off frequency for each mode). Thus
the energy velocity $V$ is written in terms of $\omega $ and
$k_{z^{\prime }}$ as
\begin{equation}
V=\frac{c^{2}k_{z^{\prime }}}{\omega }=\frac{\partial \omega }{\partial
k_{z^{\prime }}}<c  \label{eq34}
\end{equation}
As could be expected, the energy velocity coincides with the group
velocity $\partial \omega /\partial k_{z^{\prime }}$.

Notice that $\omega k_{z^{\prime }}$ equals the expression contained in the
transmitted power. So
\begin{equation}
P_{mn}=\frac{\omega _{mn}k_{z^{\prime }}}{2\,\mu _{o}}\int \psi _{mn}^{2}\,dS
\label{eq35}
\end{equation}
If the waveguide is filled with an homogeneous linear medium, then
the Eq.~(\ref{eq35}) for the transmitted power can be used in the
frame where the medium is at rest by replacing $\mu _{o}$ with the
permeability $\mu $ (the constitutive relations are only valid in
the media proper frames).

To end the study of the energy transmission let us compute the time-averaged
energy density $c^{2}T^{t^{\prime }t^{\prime }}=c^{-2}T_{t^{\prime
}t^{\prime }}$. The results to be obtained are common to TM and TE modes,
but it will be easier to compute them for TM modes. We start from
\begin{equation}
\mu _{o}\,<T_{t^{\prime }t^{\prime }}>=g^{xx}<F_{t^{\prime }x}F_{t^{\prime
}x}>+g^{yy}<F_{t^{\prime }y}F_{t^{\prime }y}>+<F_{t^{\prime }z^{\prime
}}F_{t^{\prime }z^{\prime }}>+\frac{c^{2}}{4}<F_{\lambda \rho }F^{\lambda
\rho }>
\end{equation}
The invariant $<F_{\lambda \rho }F^{\lambda \rho }>$ can be computed with
the components of the stationary field (it is invariant under a Lorentz
boost!). The field of stationary TM modes has only three independent
components. The real fields are
\begin{eqnarray}
F_{tz} &=&-\Omega _{mn}\;\psi _{mn}\sin (\Omega _{mn}\,t)  \nonumber \\
F_{xz} &=&\partial _{x}\psi _{mn}\;\cos (\Omega _{mn}\,t)  \label{eq22} \\
F_{yz} &=&\partial _{y}\psi _{mn}\;\cos (\Omega _{mn}\,t)  \nonumber
\end{eqnarray}
Therefore, the time-averaged scalar invariant $<F_{\lambda \rho }F^{\lambda
\rho }>$ is
\begin{eqnarray}
&<&F_{\lambda \rho }F^{\lambda \rho
}>=2<F_{tz}F^{tz}+F_{xz}F^{xz}+F_{yz}F^{yz}>  \label{eq23} \\
&=&-c^{-2}\Omega _{mn}^{2}\psi _{mn}^{2}+g^{xx}\,(\partial _{x}\psi
_{mn})^{2}+g^{yy}(\partial _{y}\psi _{mn})^{2}=-c^{-2}\Omega _{mn}^{2}\psi
_{mn}^{2}+|\mathbf{\nabla }\psi _{mn}|^{2}  \nonumber
\end{eqnarray}
Then
\begin{equation}
\mu _{o}\,<T_{t^{\prime }t^{\prime }}>=\frac{1}{2}\gamma
(V)^{2}V^{2}| \mathbf{\nabla }\psi |^{2}+\frac{1}{2}\Omega
^{2}\psi ^{2}-\frac{1}{4} \,\Omega ^{2}\psi
^{2}+\frac{c^{2}}{4}|\mathbf{\nabla }\psi |^{2}=\frac{1}{4} \Omega
^{2}\psi
^{2}+\frac{c^{2}}{4}\,\frac{1+\frac{V^{2}}{c^{2}}}{1-\frac{
V^{2}}{c^{2}}}\;|\mathbf{\nabla }\psi |^{2}  \label{eq36}
\end{equation}
By performing the integral in the waveguide section we obtain the energy per
unit of length:
\begin{equation}
U_{mn}=\int c^{2}T^{t^{\prime }t^{\prime }}\,dS=\frac{1}{4\mu
_{o}c^{2}}\int \left\{ \Omega _{mn}^{2}\psi
_{mn}^{2}+c^{2}\,\frac{1+\frac{V^{2}}{c^{2}}}{1-
\frac{V^{2}}{c^{2}}}\;|\mathbf{\nabla }\psi _{mn}|^{2}\right\}
\;dS=\frac{\gamma (V)^{2}\Omega _{mn}^{2}}{2\mu _{o}c^{2}}\int
\psi _{mn}^{2}\,dS \label{eq37}
\end{equation}
where we have used the result (\ref{eq29}). By comparing with Eq.
(\ref{eq35}) one gets
\begin{equation}
P_{mn}=V\,U_{mn}  \label{eq38}
\end{equation}
Once again we find the velocity $V$ in the expected role of the
energy velocity: in this case as the quotient of transmitted power
and energy per unit of length as is usually defined.\cite{jack}
Notice that $T^{z^{\prime }t^{\prime }}$ is the density of the
$z^{\prime }$-component of the momentum. Therefore, since the
energy-momentum tensor is symmetric, the former expression can
also be read as the momentum per unit of length equaled to the
energy per unit of length over $c^{2}$ times the energy velocity
(the usual relativistic relation for massive particles!). Even
though the electromagnetic field is massless its energy in the
waveguide behaves like the one of a massive field, as a
consequence of the boundary conditions imposed by the waveguide.
The same feature emerges in theories with compactified dimensions,
which impose periodic boundary conditions to the (otherwise
massless) fields.\cite{kaku}

\bigskip

\section{Appendix}

Let $\phi, \psi$ be two differentiable functions ($0$-forms). Then
one has the following identity among \textsl{volumes} ($n$-forms),
\begin{equation}
d\phi \wedge \, \ast d\psi +\phi \;d\ast d\psi=d(\phi \;\ast d\psi
) \label{eq50}
\end{equation}
which can be integrated to become
\begin{equation}
\int_S (d\phi \wedge \, \ast d\psi +\phi \;d\ast
d\psi)=\int_{\partial S}\phi \;\ast d\psi
\end{equation}
(Stokes theorem \cite{schu,flan,darl,west} has been used on the
right side). The $n$-form $d\ast d\psi$ is connected with
$\Delta\psi =-\ast d\ast d\psi$. According to Eq.~(\ref{eq40}) it
is $\ast \Delta\psi=-d\ast d\psi$ whenever $\psi$ is a $0$-form
and the space has signature $\sigma=n$. Thus we get the Green's
first identity
\begin{equation}
\int_S (d\phi\wedge \, \ast d\psi  -\phi \;\ast
\Delta\psi)=\int_{\partial S}\phi \;\ast d\psi\label{eq60}
\end{equation}

To prove Eq.~(\ref{eq29}) we will apply Green's first identity to
the case $\phi=\psi$ in the waveguide section, which is a
2-dimensional manifold with coordinates $x,y$ and signature
$\sigma=2$. For any differentiable function (0-form) $\psi $ one
has
\begin{eqnarray}
d\psi &=&\partial _{x}\psi \;dx+\ \partial _{y}\psi \;dy \nonumber
\\ \ast d\psi &=&\sqrt{g_{xx}g_{yy}}\ (\ g^{yy}\;\partial _{y}\psi
\;dx-\ g^{xx}\;\partial _{x}\psi \;dy) \label{eq49} \\ d\psi\wedge
\, \ast d\psi  &=&-\sqrt{g_{xx}g_{yy}}\ \left( g^{xx}\;(\partial
_{x}\psi )^{2}+g^{yy}\;(\partial _{y}\psi )^{2}\right) \,dx\wedge
dy \nonumber
\end{eqnarray}
As stated, the 2-form $d\psi\wedge \,\ast d\psi$ is a volume in
the 2-dimensional waveguide section $S$. It contains the surface
element $dS=\sqrt{g_{xx}g_{yy}}\,dx\,dy$. Notice that the integral
we are interested in Eq.~(\ref{eq29}) is
\begin{equation}
\int |\mathbf{\nabla }\psi |^{2}\ dS=-\int_{S}d\psi \wedge \,\ast
d\psi
\end{equation}
So, we can compute it by using Eq.~(\ref{eq60}) for $\phi=\psi$.
Since we are interested in functions $\psi$ that vanish on the
boundary (TM modes), or their normal derivatives vanish on the
boundary (TE modes) (i.e., $\ast d\psi$ restricted to the boundary
is null; see Eq.~(\ref{eq62})), then the right side of Eq.
(\ref{eq60}) is zero in these cases. Besides, it is $\ast
1=\sqrt{g_{xx}g_{yy}}\ dx\wedge dy$, which will be used to compute
the Hodge star of the $0$-form $\Delta\psi$. Moreover, we are
working with functions accomplishing the eigenvalue equation
$\Delta\psi=-\Omega^2\, c^{-2}\, \psi$; then it is $\ast
\Delta\psi=-\Omega^2\, c^{-2}\, \psi\, \sqrt{g_{xx}g_{yy}}\
dx\wedge dy$. Thus the result is
\begin{equation}
\int |\mathbf{\nabla }\psi |^{2}dS=-\int_{S}d\psi\wedge \,\ast
d\psi
 =-\int_{S}\psi\ \ast\Delta\psi=\int_{S}c^{-2}\Omega ^{2}\psi
^{2}\ \sqrt{g_{xx}g_{yy}}\ dx\wedge dy=\int c^{-2}\Omega ^{2}\
\psi ^{2}dS
\end{equation}

\bigskip

It should be emphasized that expressions like the ones in
Eq.~(\ref{eq49}) only depend on the \emph{normalized} basis of
forms and vectors:
\begin{equation}
\mathbf{e}^{\hat{x}}=\sqrt{g_{xx}}\ dx,\ \ \ \ \ \ \ \
\mathbf{e}^{\hat{y}}= \sqrt{g_{yy}}\ dy
\end{equation}
\begin{equation}
\mathbf{e}_{\hat{x}}=\sqrt{g^{xx}}\ \frac{\partial }{\partial x},\
\ \ \ \ \ \ \ \mathbf{e}_{\hat{y}}=\sqrt{g^{yy}}\ \frac{\partial
}{\partial y}
\end{equation}


\begin{thebibliography}{99}
\bibitem{footnote} Green's first identity can be stated as
$\int_S (\phi\,\Delta\psi+\nabla\phi\cdot\nabla\psi)\,
dS=\int_{\partial S} \phi\, \frac{\partial\psi}{\partial n}\,
d\ell$, where $d\ell$ is the length element on the boundary
$\partial S$ and $\partial/\partial n$ is the normal derivative.
See the Appendix for a proof using exterior calculus.
\end{thebibliography}
\end{document}